\documentclass[a4paper]{jpconf}

\usepackage{graphicx}
\newcommand{\be}{\begin{equation}}
\newcommand{\ee}{\end{equation}}  	 
\newcommand{\ba}{\begin{eqnarray}}      
\newcommand{\ea}{\end{eqnarray}}	 
\newcommand{\bite}{\begin{itemize}}
\newcommand{\eite}{\end{itemize}}
\begin{document}
\title{Charmed hadron spectroscopy on the lattice for $N_f=2+1$ flavours }

\author{Gunnar Bali, Sara Collins and Paula P\'erez-Rubio\footnote{Speaker.}}

\address{Institut f\"ur Theoretische Physik, Universt\"at Regensburg, D-93040 Regensburg, Germany }

\ead{gunnar.bali@physik.uni-r.de, sara.collins@physik.uni-r.de, paula.perez-rubio@physik.uni-r.de}

\begin{abstract}
We study the spectra of charmonia, charmed mesons, singly and
doubly charmed baryons using Lattice QCD, with 2+1 flavours of fermions. 
In the case of mesons, we include higher spin states, 
while for baryons, both positive and negative parity channels 
were investigated. By means of the variational method, 
we were able to extract a clean signal from the correlation functions and 
information about the excited states. 
\end{abstract}

\vspace{-2.7em}

\section{Introduction}

\vspace{0.5em}

In recent years the spectroscopy of hadrons containing 
charm quarks has received special attention. Concerning 
charmonium-like states, since 2002, new resonances have 
been found that are unlikely to be  conventional $\bar c c$ 
states (see ref, \cite{Brambilla:2010cs} for  
a summary of quarkonium states). The 
triggering point was the discovery of the $X(3872)$ 
resonance by the Belle collaboration \cite{Choi:2003ue},
later confirmed by BaBar, CDF and D0, whose mass
lies close to the $D\bar D^*$ threshold. It
has been widely studied and no consensus  on its inner 
structure has been reached so far. After this observation, 
a plethora of puzzling 
chamonium-like states emerged from the experiments, the 
most interesting being the ones lying close to the open charm 
thresholds. These new states could correspond to loosely bound 
hadronic molecules, hybrid states or tetraquark states.

\vspace{0.5em}

We find a similar situation  in  open charm meson 
spectroscopy, although less states have been found. In 2002, the 
BaBar Collaboration discovered a positive parity 
scalar meson, $D_{sJ}(2317)$ \cite{Aubert:2003fg} 
which was confirmed soon after by CLEO, 
\cite{Besson:2003cp}. In the same experiment, CLEO 
observed a $1^{+}$ state, $D_{sJ}(2460)$. Both resonances 
are very narrow and lie right below the 
$D K$  and $D^*K$ thresholds, respectively. 
Recently, more  new states were 
observed, for example the 
$D^{*}_{s1}(2700)$ and $D^{*}_{sJ}(2860)$.

\vspace{0.5em}

\begin{figure}[h]
\begin{center}
\begin{minipage}{14pc}
\vspace{-0.5em}
\includegraphics[width=7pc]{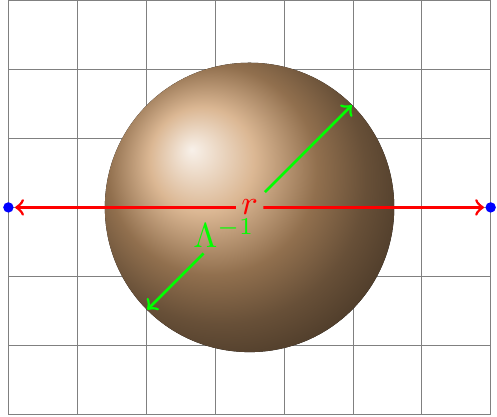}
\end{minipage}\hspace{2pc}%
\begin{minipage}{14pc}
\includegraphics[width=7pc]{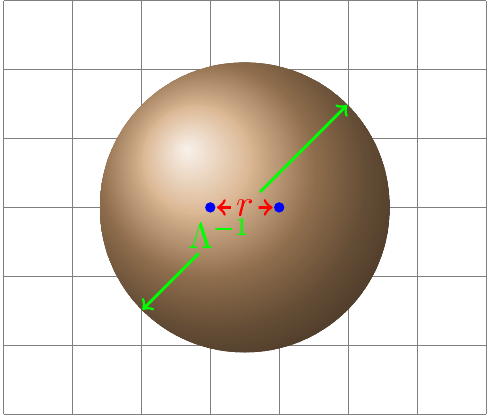}
\end{minipage} 
\caption{\label{label1}  \footnotesize{ Structure 
of a doubly charmed baryon: Charmonium-like picture (left), 
HQET picture (right).}}
\vspace{-1.5em}
\end{center}
\end{figure}

Concerning charmed baryons, there exist 17 experimentally 
 well established  singly charmed baryons 
\cite{Beringer:1900zz},   most of which have been found in 
the last decade, in the $e^+e^-$ colliders and at Fermilab. 
The charm quark is sufficiently
massive for the states to be described as a combination 
of a heavy quark and a light di-quark and their properties
can be understood in terms of Heavy Quark Effective Theory (HQET). 
Moreover, doubly-charmed baryons are also interesting, 
since they combine two scales of QCD:  the size of the two heavy quark ($QQ$)
system and  $\Lambda^{-1}$ (where $\Lambda$ is a typical  
binding energy). Two possible pictures are 
shown in Figure \ref{label1}. In the charmonium-like picture (left), 
the di-quark is formed by a heavy and a light quark. The resulting object 
will interact with the remaining heavy $Q$, as if it was a charmonium system. 
The radius of the $QQ$ system is much larger than  $\Lambda^{-1}$. 
In the HQET  picture (right), 
the $QQ$ diquark system binds itself into the $\bar 3$ 
representation of SU(3). In that case the radius of the 
heavy-heavy diquark is smaller than  $\Lambda^{-1}$.

In the next few years, more  results are expected to
appear in currently running experiments,
e.g. Belle, BES-III and LHC and the future PANDA experiment
at the FAIR facility. In order to study these systems
from a theoretical point of view, we need a 
non perturbative approach to QCD. Lattice QCD achieves this 
through a Monte Carlo evaluation of the path integral after a 
discretisation of space time on a finite lattice 
with lattice spacing $a$. To keep systematic 
errors under control,  we need  $a^{-1}$ to be much larger than the 
relevant physical scales of the problem, the size of 
the box $L$ to be larger than the typical 
size of the hadrons and an extrapolation to the physical quark masses 
to be carried out (simulating physical msses is computationally expensive).

\section{Methods and computational details}

\vspace{0.5em}
We have employed  $N_f = 2+1$ gauge configurations generated 
using the tree level  ${\rm O}(a^2)$ Symanzik 
improved Wilson action for the  gluonic degrees of freedom. 
The fermionic action uses non-perturbatively improved 
Wilson fermions with stout links in the derivative terms 
(SLiNC  \cite{Cundy:2009yy}). The quark masses were  
first tuned  to the  SU(3)$_{\rm flavour}$ -symmetric point
where the flavour singlet mass average $m_q = (m_u + m_d + m_s)/3$ 
takes its physical value. Then, the different 
quark masses are varied while keeping the singlet 
quark mass  fixed \cite{Bietenholz:2010jr,Bietenholz:2011qq}. 
At present, there is only one $\beta$ value available, $\beta= 5.5$, 
corresponding to $a \sim  0.0795$ fm.  

\begin{table}[ht!]
\begin{center}
\begin{tabular}{|c|c|c|c|c|c|}
\hline
$\kappa_l$&  $\kappa_s$ & $a$ fm &\footnotesize{\# meas (meson)}& 
\footnotesize{\# meas (baryon)} &  $M_\pi$ (MeV)\\
\hline \hline
$0.12090$&$0.12090$
&  {$0.0795(3)$} & $941$ & $--$ &{$442$} \\
$0.12104$&$0.12062$&
{$0.0795(3)$} & $450$ & $450$ & {$348$}\\
\hline
\end{tabular}
\caption{\label{table1} Details of the configurations used so far in this study.}
\vspace{-5mm}
\end{center}
\end{table}

Let $\hat O_1$ and $\hat O_2$ be two  interpolating 
operators overlapping with the  state we are interested 
in. A  correlation function can then be  defined,
\ba
C(\hat {\mathcal O}_i, \hat {\mathcal O}_j, t) &= &
\langle\hat {\mathcal O}_j(0) \hat {\mathcal O}_i^{\dagger}(t)
  \rangle = \lim_{T\to \infty }\frac{1}{Z(T)}{\rm Tr} 
\left[e^{-(T -t)\hat H}\hat {\mathcal O}_j
e^{-t\hat H}\hat {\mathcal O}_i^\dagger \right] = \nonumber \\
&= & \sum_n \langle 0|\hat {\mathcal O}_j|n\rangle\langle n|
\hat {\mathcal O}^\dagger_i|0\rangle 
e^{-E_nt} , \quad Z(T) = {\rm e^{-T\hat H}},
\label{eq1}
\ea
with $T$ being the temporal extent of the lattice.
Every correlation function contains a tower of states. 
Since we are primarly interested in the lower lying 
states, the contributions for higher excitations can 
represent a problem if they do not die out sufficiently 
fast. The variational method \cite{Luscher:1990ck}
allows us to disentangle different states.
The idea is to choose a basis of operators $\hat O_i$  and construct 
a cross correlation matrix  $[C(t)]_{ij} = C(\hat O_i, \hat O_j,t) $. 
One then solves the generalised eigenvalue problem (GEVP), 
\be
C^{-1/2}(t_0) C(t) C^{-1/2}(t_0) v^{\alpha}(t,t_0) = 
\lambda^\alpha (t, t_0) v^{\alpha}(t, t_0).
\ee
It can be shown that the eigenvalues behave like, 
\be
\lambda^\alpha(t,t_0) \propto e^{-(t-t_0)E_\alpha}
\left[1 + O\left(e^{-\Delta E_\alpha(t-t_0)}\right)\right]
\quad \Delta E_\alpha = E_{\alpha'} - E_\alpha \textrm{  with }
\alpha' > \alpha.
\ee
For both mesons and  baryons we use a basis of three operators. 
Each operator has the same Fock structure but different 
spatial extent. The latter is achieved by applying gauge invariant
smearing  on the  quark fields.

\vspace{0.3em}

\section{Mesons with hidden and open charm} 
\vspace{0.5em}

\begin{center}
\begin{figure}[h]
\begin{minipage}{18pc}
\includegraphics[width=18pc]{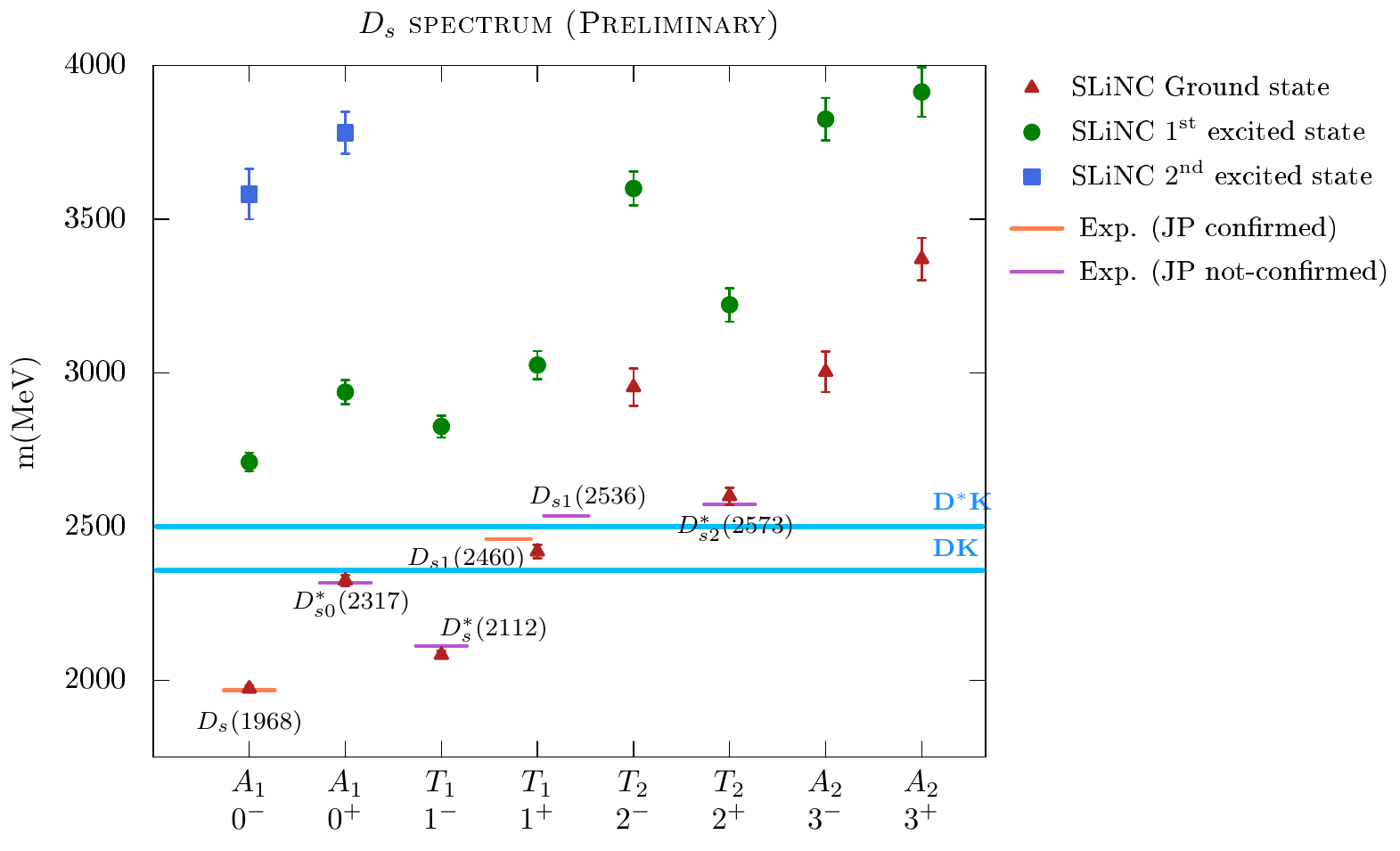}
\end{minipage}\hspace{2pc}%
\begin{minipage}{18pc}
\includegraphics[width=18pc]{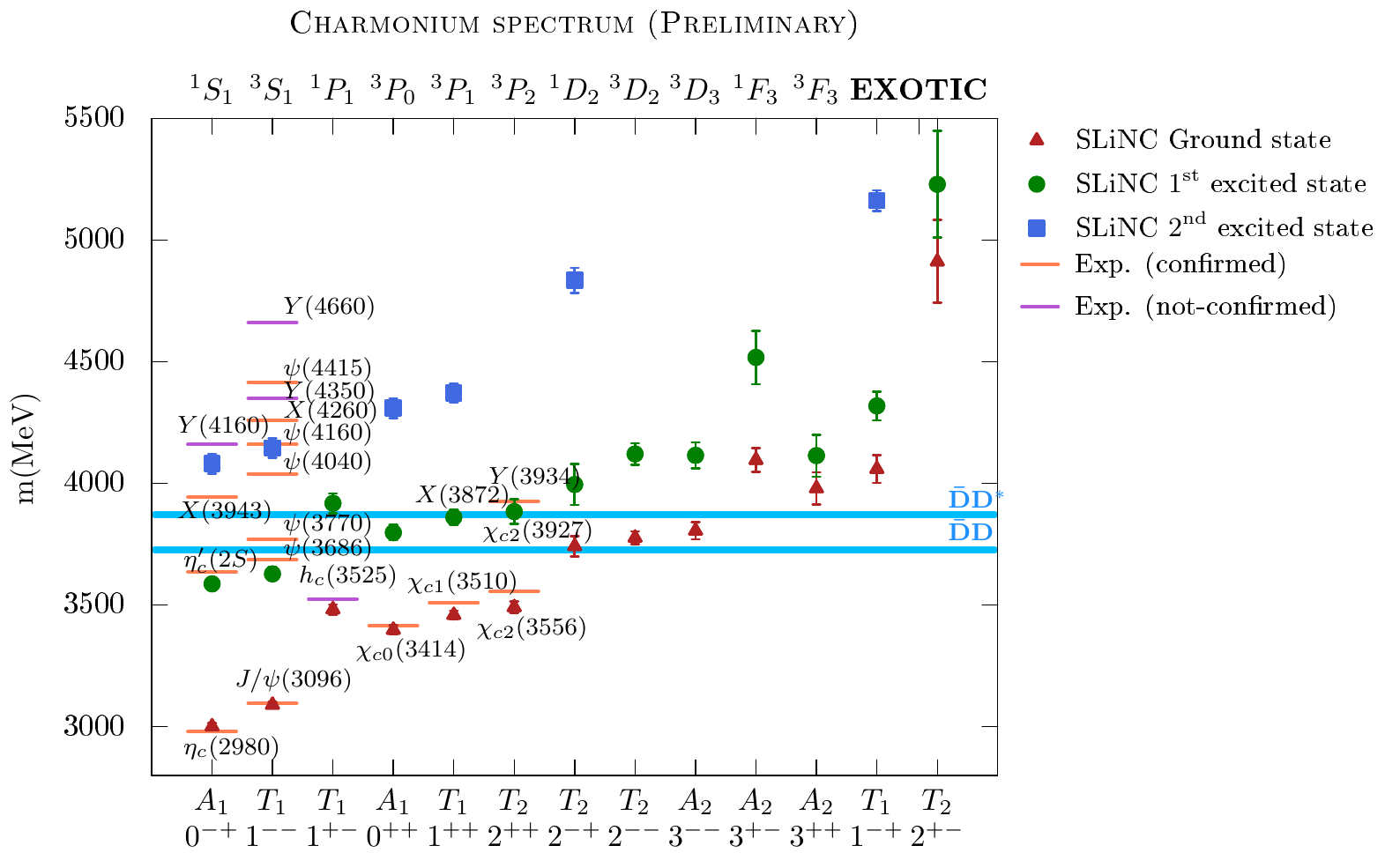}
\end{minipage} 
\caption{\label{label2}  \footnotesize{$D_s$ (left) and 
charmonium (right) spectra from $N_f = 2+1$ configurations.
The  points  are results from  the flavour symmetric
 ensemble. Error bars are statistical only.}}
\vspace{-2.7em}
\end{figure}
\end{center}

To compute the meson spectra, we used a 
subset of the operators given in Ref \cite{Dudek:2007wv}, which
have the general form,  
\be
O(x) =  \bar q_1(x) \Gamma D q_2(x),
\ee
where the indices denote  flavour, $\Gamma$ is a combination 
of gamma matrices and $D$ represents a covariant derivative operator, 
containing zero, one or two  derivatives depending on the spatial 
angular momentum of the particle under consideration (we have 
explored states with $L\le2$). The results obtained for the spectra of the 
$D_s$ and charmonium systems are shown in Figure \ref{label2}, for the 
flavour symmetric ensemble $(\kappa_l=\kappa_s)$.
We could resolve the first excited state as well as the ground state 
for all operators. In most  charmonium states and 
for two $D_s$ operators, the second excited state was also extracted. 
We can see that the experimental spectra are qualitatively reproduced. 
Before making a more detailed comparison,
results at different light quark masses and volumes are required.
This work is in progress. 
The good signals obtained for radially excited and non-zero 
angular momentum states gives us confidence in our methods.
For the  $D_s$ system, we will study the mixing of the 
$0^+$ and $1^+$ states  with the $DK$ and $DK^*$ molecules, 
respectively. Analogously, we will study molecule 
mixing for the charmonium-like $1^{++}$ state, lying close 
to the $(D^0)^*\bar D^0$ threshold .

\section{Charmed baryons} 
\vspace{0.4em}
\subsection{Interpolating operators}
\vspace{0.4em}

\begin{figure}[ht!]
\begin{center}
\includegraphics[width=13pc]{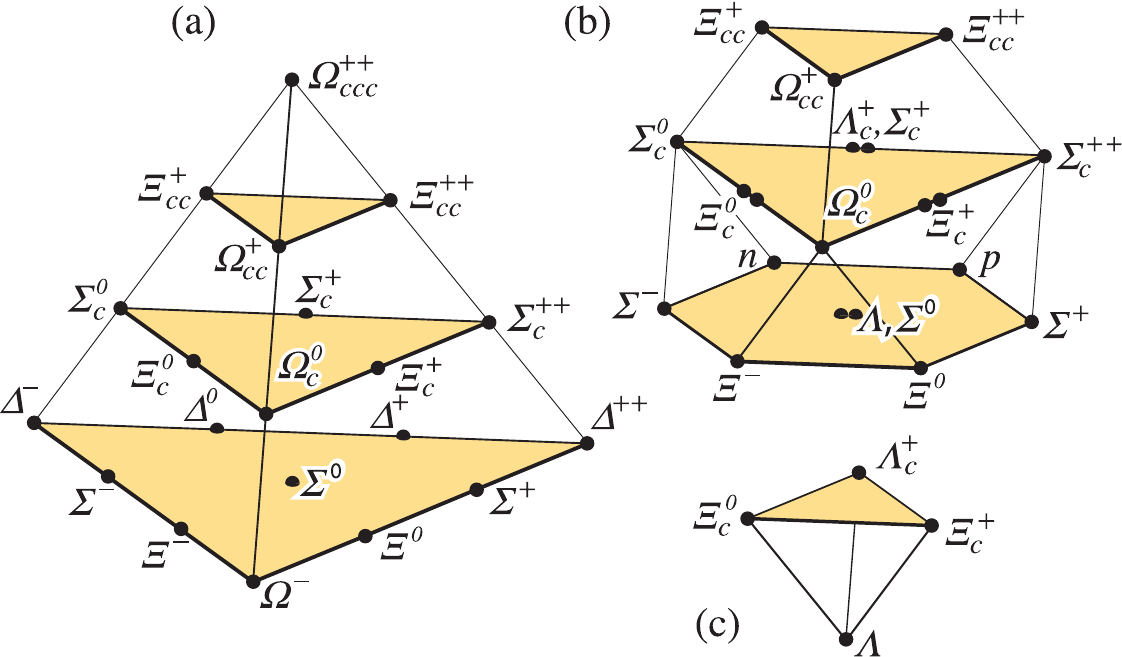}
\caption{\label{irreps}  \footnotesize{ SU(4) multiplets 
of baryons containing $u,d,s,c$ quarks. a) Totally symmetric 
20-plet with the SU(3) decuplet in the lowest level.
b) 20-plet with the SU(3) octet in the lowest level.
c) Totally antisymmetric quadruplet ($\bar 4$). 
The last one can only occur for spatial angular 
momentum with odd values. Figure  from 
\cite{Beringer:1900zz}. }}
\end{center}
\vspace{-2em}
\end{figure}

Even though flavour symmetry is not an exact symmetry of nature, 
 SU(2)$_{\rm flavour}$ is reasonably well 
respected. If we include the 
strange quark,  since  $m_s - m_u <  \Lambda_{\rm QCD}$, 
we still observe an approximate flavour symmetry 
(SU(3)$_{\rm flavour}$) in the baryon spectrum. 
Pushing things further, 
when including the charm quark, flavour symmetry is not
a good symmetry of the system. 
In principle, it is not clear that the 
interpolating operators falling into 
the irreducible representations of SU(4)$_{\rm flavour}$, 
cf. Figure \ref{irreps}  
have significant overlaps with the actual baryon states. 
However, empirically, we have found that this is indeed
the case. 
\vspace{0.3em}

Alternatively, following HQET, we can use 
operators that describe a heavy baryon as 
a light di-quark in the presence of a 
heavy quark. They are listed in 
Table \ref{table3}. Comparing 
results from the two  sets of 
operators can help us understand 
the inner structure of the baryons.

\begin{table}[ht!]
{\footnotesize
\begin{center}
\setlength{\tabcolsep}{0.7mm}
\renewcommand{\arraystretch}{1.2}
\begin{tabular}{|rcr|c|c|c|c|}
\multicolumn{5}{c}{{\sc Singly charmed }} 
&\multicolumn{1}{c}{{$J^P= \frac12^+$}}
 &\multicolumn{1}{c}{{$J^P= \frac32^+$}}\\ 
\hline
  $\bf (S)$& $\bf (I)$& $\bf s_d^{\pi_d}$ & $\bf (qq)Q$ & $\mathcal O$  &\textbf{Name}&\textbf{Name}\\
\hline
\hline
 $(0)$ & $(0)$                  & $(0)^+$& $(ud)c$ & ${\mathcal O}_5=\epsilon_{abc}(u^{aT}C\gamma_5d^b)c^c  $&$\Lambda_c$ &  \\
 $(0)$ & $(1)$                  & $(1)^+$& $(uu)c$ & ${\mathcal O}_\mu=\epsilon_{abc}(u^{aT}C\gamma_\mu u^b)c^c $&$\Sigma_c$  &$\Sigma^*_c$   \\
 $(-1)$& $\left(\frac12\right)$ & $(0)^+$& $(us)c$ & ${\mathcal O}_5=\epsilon_{abc}(u^{aT}C\gamma_5 s^b)c^c $&$\Xi_c$     &      \\
 $(-1)$& $\left(\frac12\right)$ & $(1)^+$& $(us)c$ & ${\mathcal O}'_\mu=\epsilon_{abc}(u^{aT}C\gamma_\mu s^b)c^c $&$\Xi'_c$    &$\Xi^*_c$     \\
 $(-2)$& $(0)$                  & $(1)^+$& $(ss)c$ & ${\mathcal O}_\mu=\epsilon_{abc}(s^{aT}C\gamma_\mu s^b)c^c $&$\Omega_c$  &$\Omega^*_c$   \\
\hline
\multicolumn{6}{c}{}\\
\multicolumn{5}{c}{\sc  Doubly charmed}&\multicolumn{1}{c}{{$J^P= \frac12^+$}}
 &\multicolumn{1}{c}{{$J^P= \frac32^+$}}\\ 
\hline
  $\bf (S)$& $\bf (I)$& $\bf s_d^{\pi_d}$ & $\bf (QQ)q$ & $\mathcal O$  &\textbf{Name}&\textbf{Name}\\
\hline
\hline
 $(0)$ & $(0)$                  & $(1)^+$& $(cc)u$ & ${\mathcal O}_\mu=\epsilon_{abc}(c^{aT}C\gamma_\mu c^b)u^c $&$\Xi_{cc}$    &$\Xi^*_{cc}$     \\
 $(-1)$& $\left(\frac12\right)$ & $(1)^+$& $(cc)s$ & ${\mathcal O}_\mu=\epsilon_{abc}(c^{aT}C\gamma_\mu c^b)s^c $&$\Omega_{cc}$ &$\Omega^*_{cc}$  \\
\hline
\end{tabular}
\caption{\label{table3} \footnotesize Summary of quantum numbers of heavy baryons 
and interpolating operators following  the HQET approach. 
Some of these operators were  suggested in 
\cite{Bowler:1996ws} for lattice calculations.
}
\vspace{-2.7em}
\end{center}
}
\end{table}


\subsection{Results}
\vspace{0.3em}

\begin{center}
\begin{figure}[h]
\begin{minipage}{17pc}
\includegraphics[width=16pc]{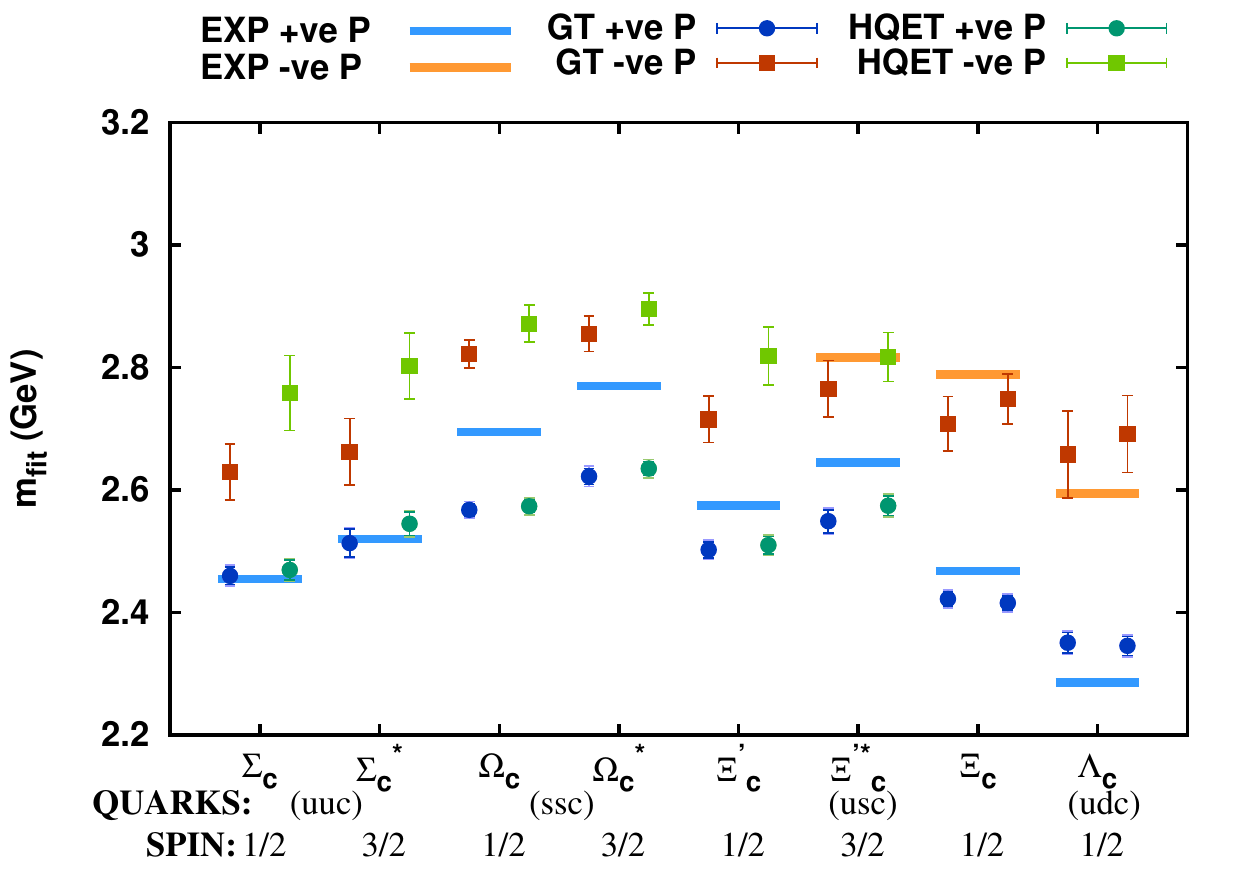}
\end{minipage}\hspace{2pc}%
\begin{minipage}{17pc}
\includegraphics[width=16pc]{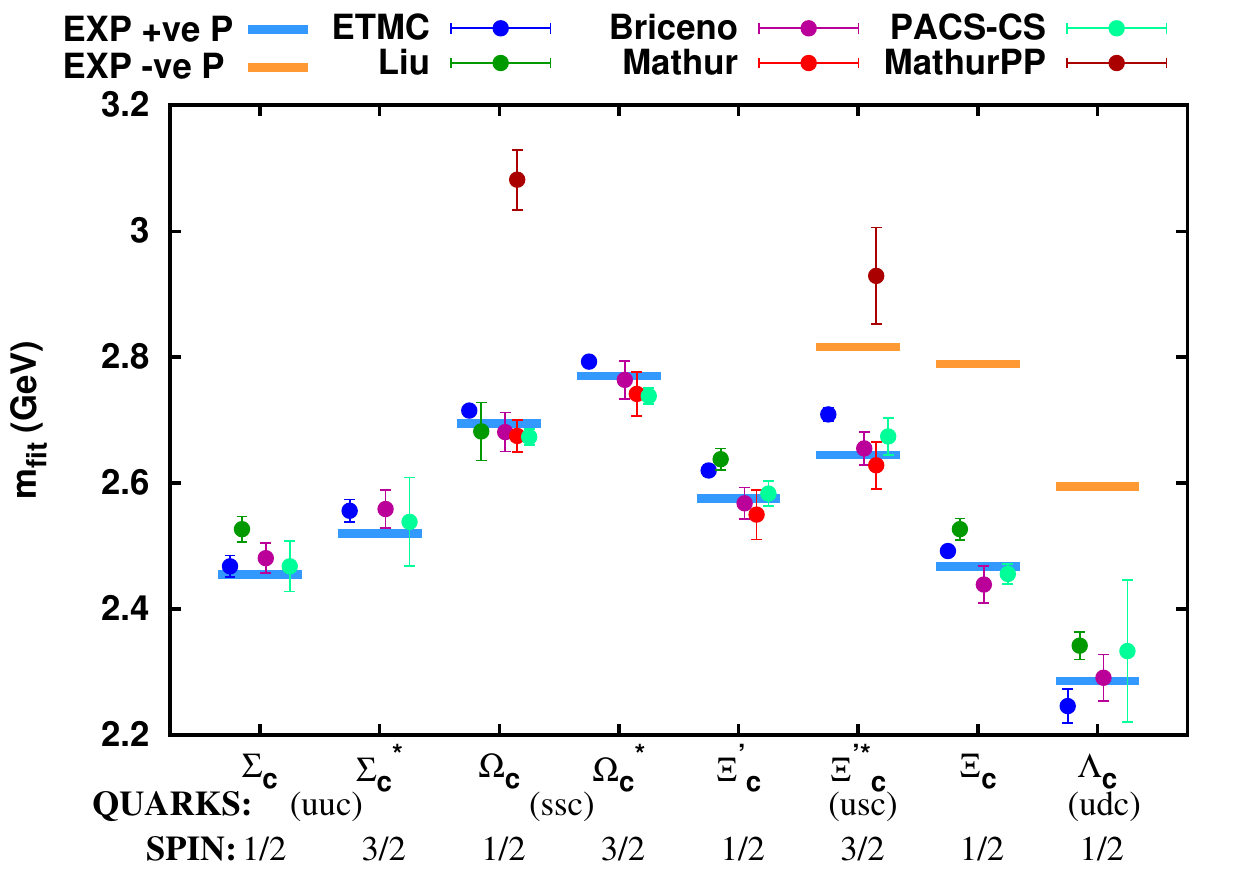}
\end{minipage}
\vspace{2pc} 

\begin{minipage}{17pc}
\includegraphics[width=16pc]{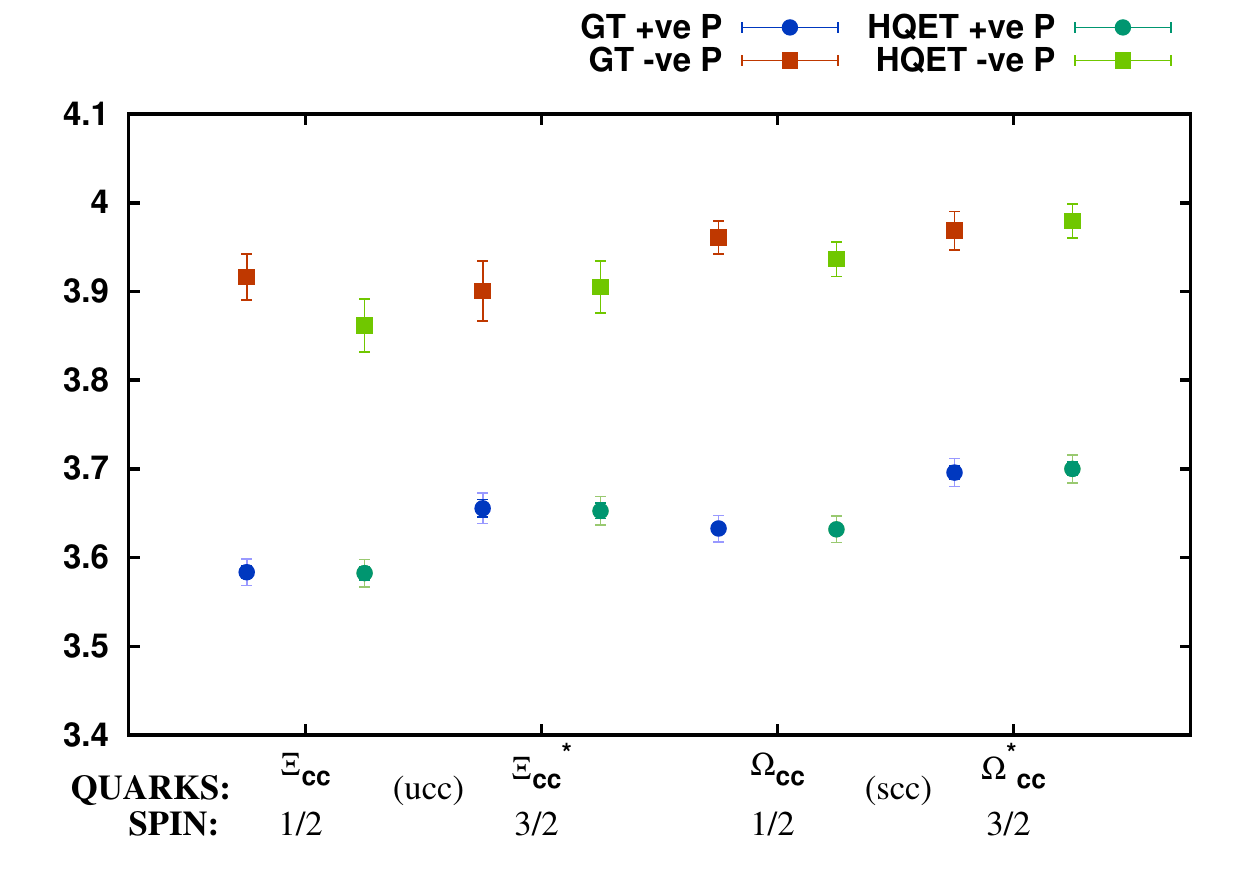}
\end{minipage}\hspace{2pc}%
\begin{minipage}{17pc}
\includegraphics[width=16pc]{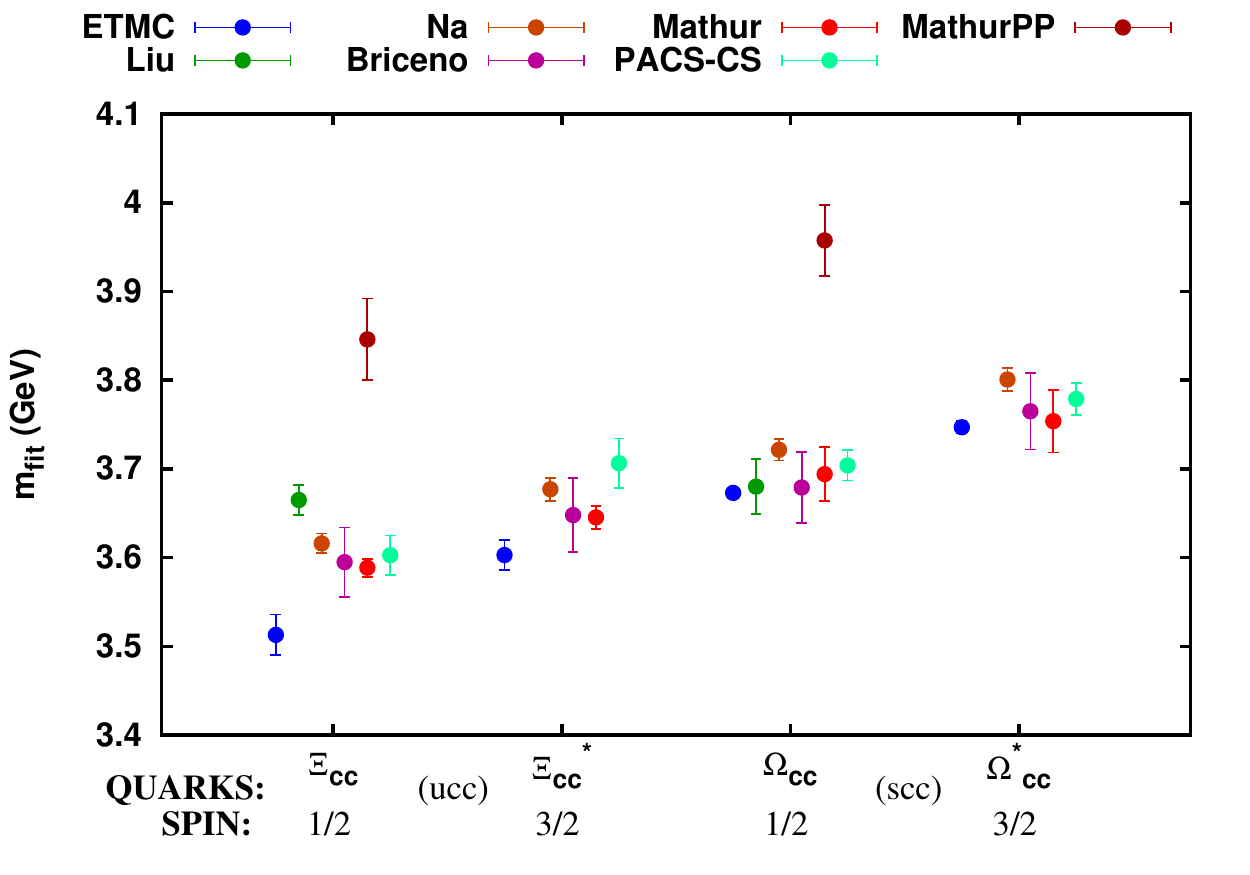}
\end{minipage} 
\caption{\label{label3}  \footnotesize{
Singly charmed (top) and doubly charmed (bottom) low 
lying spectrum. On the left hand size, results from the
SLiNC configurations are shown ($M_\pi = 348 $MeV). Errors are statistical
only. On the right hand size, a summary of 
lattice results is presented. 
}}
\vspace{-0.5em}
    \end{figure}
\end{center}

The singly and doubly charmed baryon 
spectra  for the $\kappa_l = 0.12104, 
 \kappa_s = 0.12062$ ensemble are shown in Figure \ref{label3}. 
On the left hand side, we present our preliminary singly 
(above) and doubly (below)  charmed 
baryon spectra for the two bases 
of operators chosen, including positive and negative parity channels. 
As was the case for meson spectroscopy, the simulations are 
still at an early stage. 
In our results, we can see that the 
mass differences between baryons containing  
$u,d$ quarks and the ones with  
$s$ quarks are smaller than the experimental 
values.  This is to be expected
as we are far from the physical point in terms of the masses of the
quarks: the singlet quark $m_q$ is tuned to the physical value which 
means that the $u$, $d$ and $s$ quark masses are respectively heavier 
and lighter than their physical values.
 On the right hand side, we can see a
summary of lattice results for the singly and 
doubly charmed spectra, with different systematics~\cite{Na:2007pv, 
Liu:2009jc, Alexandrou:2012xk, Briceno:2012wt, Basak:2012py, 
Namekawa:2012mp}. We can see that, overall, lattice results agree
with experiment.

\vspace{-2.8em}

\section{Acknowledgements}
\vspace{0.3em}

The numerical calculations were performed on the 
SGI Altix ICE machines at HLRN (Berlin-
Hannover, Germany). We have made use of the Chroma 
software package \cite{Edwards:2004sx}  for some
of the analysis. This work was supported by the EU ITN STRONGnet and the 
DFG SFB/TR 55. 
\vspace{0.3em}

\section{References}
\vspace{0.4em}
\bibliography{iopart-num}

\end{document}